\documentclass[12pt]{article}
\usepackage{epsf}

\begin{document}

\newcommand{\gtrsim}{\mathop{}_{\textstyle \sim}^{\textstyle >}}
\newcommand{\lesssim}{\mathop{}_{\textstyle \sim}^{\textstyle <} }

\newcommand{\rem}[1]{{\bf #1}}

\renewcommand{\thefootnote}{\fnsymbol{footnote}}
\setcounter{footnote}{0}
\begin{titlepage}

\def\thefootnote{\fnsymbol{footnote}}

\begin{center}

\hfill UT-08-24\\
\hfill TU-822\\
\hfill IPMU 08-0040\\
\hfill July, 2008\\

\vskip .75in

{\Large \bf 
Mass Measurement of the Decaying Bino at the LHC
}

\vskip .75in

{\large
$^{(a)}$Shoji Asai, $^{(a)}$Yuya Azuma, $^{(b)}$Osamu Jinnouchi, $^{(c,d)}$Takeo Moroi, 
$^{(a)}$Satoshi Shirai and $^{(a,d)}$T.T. Yanagida
}

\vskip 0.25in

{\em $^{(a)}$Department of Physics, University of Tokyo,
Tokyo 113-0033, Japan
}

\vskip 0.1in

{\em $^{(b)}$KEK, High Energy Accelerator Research Organization,
 1-1 Oho, Tsukuba 305-0801, Japan
}

\vskip 0.1in

{\em $^{(c)}$Department of Physics, Tohoku University,
Sendai 980-8578, Japan}

\vskip 0.1in

{\em $^{(d)}$Institute for the Physics and Mathematics of the Universe,\\
University of Tokyo, Chiba 277-8568, Japan}

\end{center}
\vskip .5in

\begin{abstract}

  In some class of supersymmetric (SUSY) models, the neutral Wino
  becomes the lightest superparticle and the Bino decays into the Wino
  and standard-model particles.  In such models, we show that the
  measurement of the Bino mass is possible if the short charged tracks
  (with the length of $O(10\ {\rm cm})$) can be identified as a signal
  of the charged-Wino production.  We pay particular attention to the
  anomaly-mediated SUSY-breaking (AMSB) model with a generic form of
  K\"ahler potential, in which only the gauginos are kinematically
  accessible superparticles to the LHC, and discuss the implication of
  the Bino mass measurement for the test of the AMSB model.

\end{abstract}

\end{titlepage}

\renewcommand{\thepage}{\arabic{page}}
\setcounter{page}{1}
\renewcommand{\thefootnote}{\#\arabic{footnote}}
\setcounter{footnote}{0}

\section{Introduction}

One of the strong motivations of low-energy supersymmetry (SUSY) is to
explain the origin of the dark matter of the universe.  With the
assumption of $R$-parity conservation, the lightest superparticle
(LSP) becomes stable and  is a good candidate of the dark matter
if it is weakly interacting.  The major candidates of the LSP are
the Bino $\tilde{B}$ (or, more precisely, the Bino-like lightest
neutralino) and the gravitino.

The anomaly-mediated SUSY breaking (AMSB) model \cite{Giudice:1998xp,
  Randall:1998uk}, which is a well-motivated scenario of SUSY
breaking, provides another candidate of the LSP, which is the neutral
Wino.  The Wino becomes the lightest gaugino if we adopt the mass
spectrum of the minimal anomaly-mediation.  With a generic form of the
K\"ahler potential, which is adopted in our analysis, all the
sfermions become as heavy as $O(10\ {\rm TeV})$ while the gauginos
acquire masses of $O(100\ {\rm GeV})$, and the mass spectrum of the
gauginos is modified from the prediction of the minimal case
\cite{Gherghetta:1999sw, Ibe:2006de}; importantly, even in such a
model, the neutral Wino $\tilde{W}^0$ becomes the LSP in most of the
parameter space.  Even though the thermal relic abundance of
$\tilde{W}^0$ is too small to realize the LSP dark matter, the
scenario is cosmologically viable because $\tilde{W}^0$ can be the
dark matter if it is non-thermally produced \cite{Moroi:1999zb}.  The
AMSB model with a generic form of the K\"ahler potential, which is
called the AMSB model hereafter, is a simple and natural scenario of
SUSY breaking; it is easily realized if there is no singlet field in
the SUSY breaking sector \cite{Giudice:1998xp, Wells:2004di}.

In the AMSB model with heavy sfermions, the gauginos are the primary
targets at the LHC.  In addition, since there exists a non-trivial
relation among the gaugino masses, a crucial test of the AMSB model
becomes possible once all gaugino masses in the SUSY standard model
are experimentally determined.  
In our previous studies \cite{Asai:2007sw, Asai:2008sk}, it
has been shown that properties (in particular, the masses) of the
gluino and the Wino can be studied at the LHC if the gluino is lighter
than $\sim 1\ {\rm TeV}$.  On the contrary, the Bino mass is thought
to be hardly measured at the LHC in most of the parameter space.

In this letter, we propose a new procedure to measure the Bino mass at
the LHC, using the fact that the momentum information of the charged
Wino $\tilde{W}^\pm$ from the decay of $\tilde{B}$ can be obtained if
the track of $\tilde{W}^\pm$ is observed by inner detectors.  
We pay particular attention to the measurement of the Bino mass, 
taking account of the  most recent studies of the performances of the
ATLAS detector.	We will
show that the Bino mass can be determined with the accuracy of $\sim
10 - 20\ {\rm GeV}$.  We also discuss the implication of the measurement
of the Bino mass to the test of the AMSB model.

\section{Model}

First, we summarize the model adopted  in our analysis.  Even though
there exist various possibilities to realize the Wino LSP scenario, we
concentrate on the AMSB model with a generic form of K\"ahler
potential.  Such a model has several important features.  First of
all, sfermion and Higgsino masses are not suppressed compared to the
gravitino mass, while the gaugino masses are
generated by the anomaly-mediation contributions and hence are one-loop
suppressed.  Consequently, only gauginos are within the kinematical
reach of the LHC.  In addition, once the Higgsinos become as heavy as
the gravitino, radiative corrections from the Higgs-Higgsino loop
diagrams modify the simple anomaly-mediation relation among the
gaugino masses \cite{Gherghetta:1999sw,Ibe:2006de}.  In the present
framework, the gaugino masses at the scale of the sfermion masses
$m_{\tilde{f}}$, which is assumed to be comparable to the Higgsino
mass $\mu_H$, are given by
\begin{eqnarray}
 M_1 &=& \frac{g_1^{2}}{16 \pi^{2}} \left( 11 m_{3/2} + L \right),
  \label{M1}
  \\
 M_2 &=& \frac{g_2^{2}}{16 \pi^{2}} \left( m_{3/2} + L \right),
  \\
 M_3 &=& \frac{g_3^{2}}{16 \pi^{2}} \left( -3 m_{3/2} \right),
  \label{M3}
\end{eqnarray}
where $g_1$, $g_2$, and $g_3$ are gauge coupling constants of
$U(1)_Y$, $SU(2)_L$, and $SU(3)_C$ gauge groups, respectively, and
$m_{3/2}$ is the gravitino mass.  (Hereafter, we use a convention
such that $m_{3/2}$ is real and positive.)  In addition, $L$ is a
complex parameter which parametrizes the Higgs-Higgsino loop
contributions. (Notice that $|L|$ is expected to be of the order of
the gravitino mass.) 

With the gaugino masses obtained from Eqs.\ (\ref{M1}) $-$ (\ref{M3}),
there exists a non-trivial constraint among gaugino masses.  Indeed,
approximating the on-shell gaugino masses by $M_1$, $M_2$, and
$M_3$ given in Eqs.\ (\ref{M1}) $-$ (\ref{M3}), we can find a
constraint \cite{Asai:2007sw}
\begin{eqnarray}
  \left| \frac{10 g_1^{2}}{3g_3^{2}} m_{\tilde{g}}
    - \frac{g_1^{2}}{g_2^{2}} m_{\tilde{W}} \right|
  \lesssim m_{\tilde{B}} \lesssim
  \frac{10 g_1^{2}}{3g_3^{2}} m_{\tilde{g}}
    + \frac{g_1^{2}}{g_2^{2}} m_{\tilde{W}},
    \label{BoundOnM1}
\end{eqnarray}
where $m_{\tilde{B}}$, $m_{\tilde{W}}$, and $m_{\tilde{g}}$ are
physical masses of the Bino, Wino, and gluino, respectively.  (In Eq.\
(\ref{BoundOnM1}), the mass difference between $\tilde{W}^\pm$ and
$\tilde{W}^0$, which is expected to be $\sim 155-170\ {\rm MeV}$,
is neglected.)  Even with the renormalization group
effect below the scale of $m_{\tilde{f}}$ (which will be included in
our numerical analyses), the Bino mass is constrained in the window
calculated with the  gluino and Wino masses.  This fact will
be used to test the AMSB model using experimental measurements of the
gaugino masses.

In the AMSB model with heavy sfermions, SUSY events at the LHC are
mostly from a pair production of the gluinos.
The produced gluinos cascade down to
lighter gauginos;
$\tilde{g}\rightarrow\tilde{B}q\bar{q}$ or
$\tilde{g}\rightarrow\tilde{W}q\bar{q}$.\footnote
{In this letter, we neglect two-body decay modes of the gluino
  ($\tilde{g}\rightarrow \tilde{B}g$ and $\tilde{g}\rightarrow
  \tilde{W}g$).  This can be justified by tuning the sfermion masses
  \cite{Toharia:2005gm}.  }
The branching ratios of these processes depend on sfermion masses, and
hence are free parameters.  Furthermore, once $\tilde{B}$ is produced,
it decays into charged or neutral Wino and standard-model particles.
In most of the parameter space where the Higgsino is as heavy as
sfermions, the Bino dominantly decays as
$\tilde{B}\rightarrow\tilde{W}^\pm W^\mp$ and $\tilde{W}^0 h$ (with
$h$ being the standard-model-like Higgs boson).  When $M_1$ and $M_2$
are real and have the same sign, which is the case in our numerical
analysis, the decay rates of those processes are given by
\begin{eqnarray}
  \Gamma_{\tilde{B}\rightarrow\tilde{W}^+ W^-} &=& 
  \frac{\beta_{\tilde{W}^\pm W^\mp} \kappa^2}{32\pi} 
  m_{\tilde{B}} 
  \left( 1 + \frac{m_{\tilde{W}}}{m_{\tilde{B}}} \right)^2
  \left[ 1 + \frac{2m_W^2}{(m_{\tilde{B}}-m_{\tilde{W}})^2} \right]
  \left[ 1 - \frac{m_W^2}{(m_{\tilde{B}}+m_{\tilde{W}})^2} \right],
  \\
  \Gamma_{\tilde{B}\rightarrow\tilde{W}^0 h} &=& 
  \frac{\beta_{\tilde{W}^0 h} \kappa^2}{32\pi} 
  m_{\tilde{B}} 
  \left[ \left( 1 - \frac{m_{\tilde{W}}}{m_{\tilde{B}}} \right)^2 
    - \frac{m_h^2}{m_{\tilde{B}}^2} \right],
\end{eqnarray}
where 
\begin{eqnarray}
  \beta_{\tilde{W}^\pm W^\mp}^2 = \frac
  {m_{\tilde{B}}^4 - (m_{\tilde{W}^\pm}^2 + m_W^2) m_{\tilde{B}}^2
    + (m_{\tilde{W}^\pm}^2 - m_W^2)^2}
  {m_{\tilde{B}}^4},
\end{eqnarray}
and $\beta_{\tilde{W}^0 h}$ is obtained from the above formula by
replacing $m_{\tilde{W}^\pm}\rightarrow m_{\tilde{W}^0}$ and
$m_W\rightarrow m_h$.  In addition, $\kappa\equiv
g_1g_2v\sin\beta\cos\beta\mu_H^{-1}$, where $v\simeq 174\ {\rm GeV}$
is the total vacuum expectation value of the Higgs boson.  The Bino
may also decay as $\tilde{B}\rightarrow\tilde{W}f\bar{f}$ (with $f$
being standard-model fermions) via the diagrams with sfermion
propagators.  The three-body decay processes are, however,
significantly suppressed in the present situation because the decay
rates are suppressed by $m_{\tilde{f}}^{-4}$.

In the previous study \cite{Asai:2007sw}, the Bino mass measurement in
the AMSB model was shown to be possible at the LHC only when the
processes $\tilde{B}\rightarrow\tilde{W}f\bar{f}$ have significant
branching ratios, which is realized in a special case where some of
the sfermions are much lighter than the Higgsino.
In that case \cite{Asai:2007sw}, the momentum information of the charged Wino
was not used.
The missing $E_T$ signature was used for analysis and 
the Bino mass was obtained with the endpoint of 
the $M_{q \bar{q}}$ distribution.
In this letter, on the
contrary, we will show that the Bino mass $m_{\tilde{B}}$ can be
measured with the momentum information about the charged Wino even if
the Bino dominantly decays via the two-body processes.

We demonstrate how well we can determine the Bino mass using
the momentum information of  the charged Wino.
The following MC sample are generated;
the gaugino masses are
\begin{eqnarray}
  m_{\tilde{W}}=200\ {\rm GeV}, ~~~
  m_{\tilde{B}}=400\ {\rm GeV}, ~~~
  m_{\tilde{g}}=1\ {\rm TeV},
\end{eqnarray}
while other superparticles are assumed to acquire masses of $O(10\
{\rm TeV})$.  
The above set of the gaugino masses are realized when
$m_{3/2}=39\ {\rm TeV}$, $|L|=27\ {\rm TeV}$, and $\mbox{arg}(L)=0$.
With this choice of the mass parameters, SUSY events at the LHC are
dominantly from $pp\rightarrow\tilde{g}\tilde{g}$; the cross section
for this process is about $220\ {\rm fb}$.  
The decay branching fraction of $\tilde{g}$ are also model-dependent, 
and 50\% branching fractions are assumed;
\begin{eqnarray}
  Br (\tilde{g}\rightarrow\tilde{B}q\bar{q}) = 
  Br (\tilde{g}\rightarrow\tilde{W}q\bar{q}) = 0.5,
  ~~~
  (q=u,d,s,c).
\end{eqnarray}
For simplicity, we assume that the gluino dominantly decays into the first
and the second generation quarks; such a situation is realized when the
third-generation squarks are heavier than the first- and second-generation
squarks.  If the gluino decays as
$\tilde{g}\rightarrow\tilde{W}^-t\bar{b}$ (and $\tilde{W}^+b\bar{t}$),
a high $p_T$ lepton can be emitted from the $W$-boson produced by the
top decay.  If the lepton is mis-identified as that from the Bino
decay, such events become the background.  However, in the case of
$\tilde{g}\rightarrow\tilde{W}^-t\bar{b}$, in which the $W$-boson is
not monochromatic in the rest frame of the parent particle (i.e.,
$\tilde{g}$), we do not expect a steep edge in the invariant mass
distribution of $W^\pm l^\mp$, which is a big contrast to the
distribution obtained from $\tilde{B}\rightarrow\tilde{W}^\pm W^\mp$.
(For the decay mode $\tilde{g}\rightarrow\tilde{B}t\bar{t}$, see the
later discussion.)  A detailed analysis including the above background
will be given elsewhere \cite{AJMSY}.

\section{Bino Mass Measurement}

Although the large  missing $E_T$ is  characteristic  signature of  
the conventional  SUSY events at the LHC, 
the charged Wino tracks are promising signature of the AMSB model.
The neutral Wino is the LSP and  the charged Wino is almost degenerate in mass with the LSP in the present model.
In most of the parameter space, the decay length of the
charged Wino becomes $c\tau\simeq 5\ {\rm cm}$ \cite{Feng:1999fu}.
A large number of the charged Winos can travel through the pixel detector 
and the semi-conductor tracker (SCT)  of the ATLAS detector before their decays.  
Some part of Wino can reach into the transition radiation tracker (TRT), which is located
outside of the SCT.
The charged Wino decays into the neutral Wino and a soft pion (or soft electron and neutrino).
The emitted pion or electron is too soft to be reconstructed in the tracking system.
Then the charged Wino track looks  disappear on the way.
This is a significant feature of the signal and clearly separated from the SM background processes \cite{AJMSY}.

Tracking performance of the ATLAS inner detector \cite{TrackReconst} is summarized in Table~1.
The charged Wino which travels until the 2nd layer of the SCT can be 
reconstructed by using the hits in the pixel, 1st and 2nd  layers of the SCT.
The charged Wino will be reconstructed perfectly, if they decay after the 3rd layer  (the transverse length of flight, $L_T > $ 443 mm). 
Good  momentum resolutions are also expected 
even only with the pixel and SCT detectors \cite{TrackReconst}.
\begin{table}[t]
\begin{center}
\begin{tabular}{|l|c|c|c|c|c|} \hline
SCT layer              & 1st & 2nd & 3rd & 4th & hit in TRT  \\ \hline
$L_{T}^{(\min)}$ (mm)  & 299 & 371 & 443 & 514 &  554        \\ \hline
Reconstruction efficiency (\%) & -- & 85 & 100 & 100 & 100  \\ \hline
\end{tabular} 
\caption{Reconstruction efficiencies of the track are summarized for the various decay positions.
SCT layer shows the final layer passing through before the decay and the $L_{T}^{(\min)}$ is distance between the layer and beam pipe.  
The transition radiation tracker (TRT) is located outside of the SCT.
Tracks are reconstructed with the pixel and the SCT layer inside the decay point. 
} \end{center} \end{table}
A lifetime of the observed charged particle can be measured with the TRT
as already pointed out in the previous study \cite{Asai:2008sk}.  
Since the lifetime of the charged Wino is insensitive to the SUSY parameters
(as far as in the AMSB model), such a lifetime measurement is the 
strong evidence that the observed long-lived charged particle is the  $\tilde{W}^\pm$
in the AMSB model.

After the charged Wino tracks are observed, 
it is also important to note that the momentum information
as well as the time-of-flight information obtained with the TRT and the EM calorimeter 
are available. 
The Wino mass can be determined with an accuracy of about $10\ \%$,
if enough samples of the charged Wino tracks are observed in the TRT and the EM calorimeter,
and if the mean beta of the charged Wino is less than 0.85.
The reconstruction of four-momenta of the charged Wino gives 
good chance to determine the Bino mass.

The procedure to measure the Bino mass is summarized in this session, 
when the Bino dominantly decays into $\tilde{B}\rightarrow\tilde{W}^\pm W^\mp$ and
$\tilde{W}^0 h$.  
Determination of  the Bino mass is to use the invariant mass distribution of $\tilde{W}^\pm +\mbox{lepton}$
system.  
In the signal events, isolated leptons are mainly from the
decay of $W^\pm$, which is produced by the decay of $\tilde{B}$.
The decay chain $\tilde{B}\rightarrow\tilde{W}^\pm W^\mp$, followed by
$W^\mp\rightarrow l^\mp\nu$ is used in the analysis.
Informations about the gaugino masses are imprinted into 
the distribution of the invariant mass of the
$\tilde{W}^\pm l^\mp$ system.
Since the charges of the lepton and the Wino are opposite in the signal event,
the combinatorial background can be reduced by the same-sign subtraction as shown in the following.
$M_{\tilde{W}^\pm l^\mp}$ is constrained in the region of 
\begin{eqnarray}
  M_{\tilde{W}^\pm l^\mp}^{\rm (min)}
  \leq  M_{\tilde{W}^\pm l^\mp} \leq
  M_{\tilde{W}^\pm l^\mp}^{\rm (max)},
\end{eqnarray}
where
\begin{eqnarray}
  M_{\tilde{W}^\pm l^\mp}^{\rm (min)2} &=& 
  \frac{1}{2}
  \left( 
    m_{\tilde{B}}^2+m_{\tilde{W}^\pm}^2-m_W^2
    - \sqrt{(m_{\tilde{B}}^2+m_{\tilde{W}^\pm}^2-m_W^2)^2 
      - 4m_{\tilde{B}}^2 m_{\tilde{W}^\pm}^2} \right),
  \\
  M_{\tilde{W}^\pm l^\mp}^{\rm (max)2} &=& 
  \frac{1}{2}
  \left( 
    m_{\tilde{B}}^2+m_{\tilde{W}^\pm}^2-m_W^2
    + \sqrt{(m_{\tilde{B}}^2+m_{\tilde{W}^\pm}^2-m_W^2)^2 
      - 4m_{\tilde{B}}^2 m_{\tilde{W}^\pm}^2} \right).
\end{eqnarray}
The endpoints of the distribution of $M_{\tilde{W}^\pm l^\mp}$  are determined by the gaugino masses. 
In particular, when the gaugino masses are much larger than $m_W$, $M_{\tilde{W}^\pm l^\mp}^{\rm
(max)}\simeq m_{\tilde{B}}$ and  the Bino mass can be obtained from the upper endpoint
of the $M_{\tilde{W}^\pm l^\mp}$ distribution.  
In the same limit, lower edge is Wino mass; $M_{\tilde{W}^\pm
l^\mp}^{\rm (min)}\simeq m_{\tilde{W}^\pm}$.

We generate the signal events by using
MadGraph/MadEvent packages
\cite{Stelzer:1994ta,Maltoni:2002qb,Alwall:2007st}, and the decays and
hadronizations of the standard model particles are treated by PYTHIA \cite{Sjostrand:2006za}.  
The produced events are fed into the PGS4 package \cite{PGS4} 
to simulate detector effects.  
The error in the measurement of the momentum of the charged Wino is not taken into account in PGS4.
The following selections are applied:
\begin{itemize}
\item[1.]  At least one isolated lepton (with $p_T>20\   {\rm GeV}$) is required,
\item[2.] At least one charged Wino (which travels
  transverse length longer than $L_T^{\rm (min)}$) whose charge is
  opposite to that of the isolated lepton.
\end{itemize}
In our analysis, we assume that the decay length of the charged Wino $c\tau = 5.13$ cm and that the reconstruction
efficiency reaches 100 \% in the case that the charged Wino travels transverse length longer than 37.1 cm.

\begin{figure}
 \centerline{\epsfxsize=0.6\textwidth\epsfbox{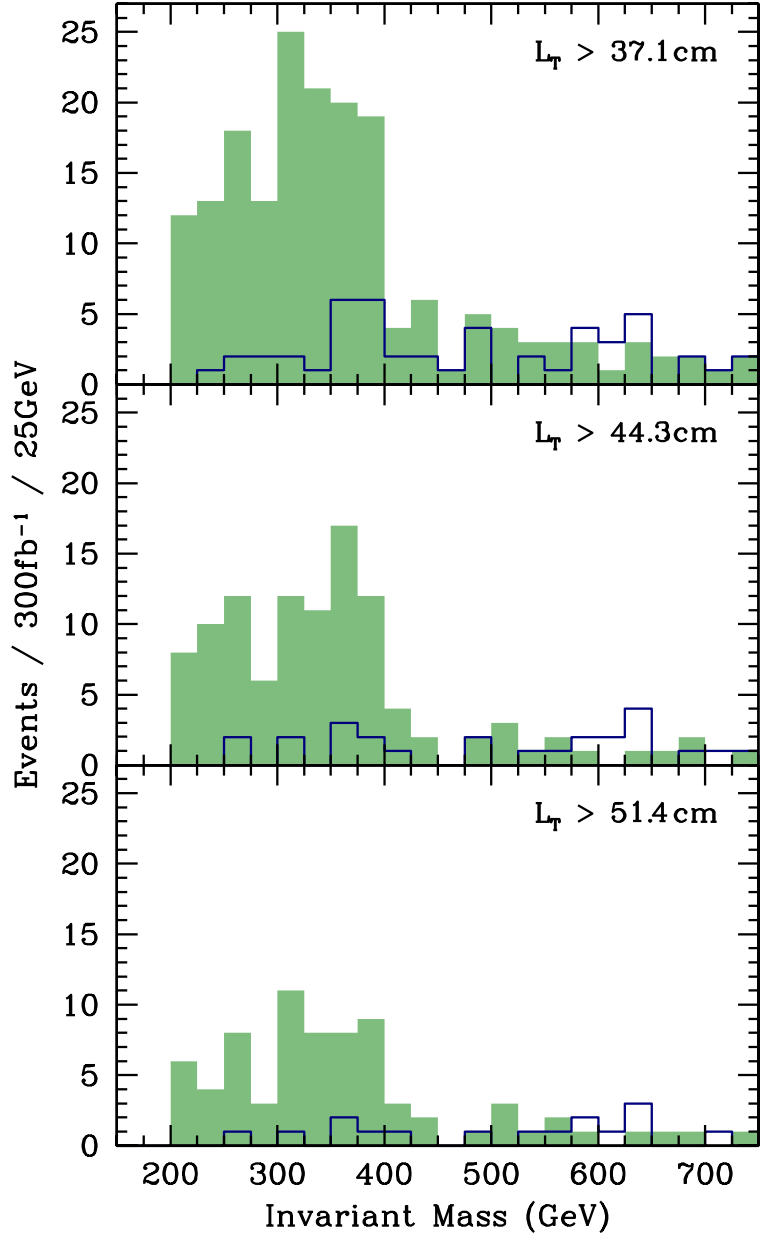}}
 \caption{\small Distribution of the opposite-sign $\tilde{W}^\pm
   l^\mp$ system (shaded histogram) and the same-sign $\tilde{W}^\pm
   l^\pm$ system (solid line) for ${\cal L}=300\ {\rm fb}^{-1}$.  Three figures 
   show the minimum transverse length of flight of  $L_T^{\rm (min)}=37.1\ {\rm cm}$, 
   $44.3\ {\rm cm}$, and $51.4\ {\rm cm} $, which are corresponding to the 2nd, 3rd, and 4th layers of SCT.}
\label{fig:minv}
\end{figure}

Fig.\ \ref{fig:minv} shows the distributions of the invariant mass of 
$\tilde{W}^\pm  l^\mp$ system for the various transverse length of flight of
$L_T^{\rm (min)}=37.1\ {\rm cm}$, $44.3\ {\rm cm}$,
and $51.4\ {\rm cm}$, which correspond to the radii of the 2nd, 3rd, and 4th layers of the SCT, respectively \cite{AtlasTdr}. 
The sharp edges at the positions of the expected endpoints are observed in these three figures.
With the present choice of gaugino masses, the expected points are $M_{\tilde{W}^\pm l^\mp}^{\rm (min)}\simeq 206\
{\rm GeV}$ and $M_{\tilde{W}^\pm l^\mp}^{\rm (max)}\simeq 389\ {\rm
  GeV}$.
An integrated luminosity of 300 ${\rm fb}^{-1}$ is assumed.
As we have mentioned, once an enough amount of the Wino tracks is identified in the TRT, 
the Wino mass can be determined with an accuracy of $\sim 10\ \%$. 
We assume the accuracy of 10\% in the following analysis and the error in the determination of
$m_{\tilde{B}}$ arising from the error in $m_{\tilde{W}^\pm}$ will be
discussed later.  
The energy of $\tilde{W}^\pm$ is estimated
from a postulated value of the Wino mass $m_{\tilde{W}^\pm}^{\rm
  (obs)}$, which is from the result of the Wino mass measurement, and
the momentum ${\bf p}_{\tilde{W}^\pm}$, as
$E_{\tilde{W}^\pm}=\sqrt{m_{\tilde{W}^\pm}^{{\rm (obs)}2}+{\bf
    p}_{\tilde{W}^\pm}^2}$, and we calculate $M_{\tilde{W}^\pm l^\mp}$ as
\begin{eqnarray}
  M_{\tilde{W}^\pm l^\mp}^2 = m_{\tilde{W}^\pm}^{{\rm (obs)}2}
  + 2\left(
    E_{\tilde{W}^\pm} E_{l^\mp} - {\bf p}_{\tilde{W}^\pm} {\bf p}_{l^\mp}
  \right),
\end{eqnarray}
where $p_{l^\mp}\equiv (E_{l^\mp},{\bf p}_{l^\mp})$ is the
four-momentum of the lepton.

There is the wrong combination of  the $\tilde{W}^\pm$ and
$l^\mp$ in the signal events.
In some events, two Binos are produced and the observed $\tilde{W}^\pm$ and $l^\mp$ are from
different Binos.  In addition, with the decay mode of
$\tilde{g}\rightarrow\tilde{B}t\bar{t}$, which is assumed to be
suppressed in our analysis, a lepton may be produced by the decay of
$W^\pm$ from the decays of $t$ and $\bar{t}$.
This lepton contributes to the combinatorial background of $M_{ \tilde{W}^\pm
 l^\mp }$  distribution.
These combinatorial backgrounds can be estimated  with the same-sign
events, because the Bino decays into $\tilde{W}^+$ and $\tilde{W}^-$ with
the same probability.  
The following two selections are applied to make the control sample 
to estimate the combinatorial background events;
\begin{itemize}
\item[1'.] There exists one isolated lepton (with $p_T>20\ {\rm GeV}$),
\item[2'.] There exists at least one charged Wino (with $L_T>L_T^{\rm
    (min)}$) whose charge is the same as that of the isolated lepton,
\end{itemize}
and calculate the invariant mass of the $\tilde{W}^\pm l^\pm$ system,
$M_{\tilde{W}^\pm l^\pm}$.  The distribution of $M_{\tilde{W}^\pm
  l^\pm}$ is superimposed in Fig.\ \ref{fig:minv}.  As one can see, the
number of the combinatorial background is much smaller than that of
the signal in the signal region (i.e., $M_{\tilde{W}^\pm l^\mp}^{\rm
  (min)} \leq M_{\tilde{W}^\pm l^\mp} \leq M_{\tilde{W}^\pm
  l^\mp}^{\rm (max)}$) 
  and the accurate edge of the endpoint can be obtained 
by subtracting the same-sign distribution.

We estimate the upper endpoint by fitting the structure with
Gaussian-smeared triangular fit (including the effect of background)
\cite{Bachacou:1999zb}:
\begin{eqnarray}
  \mbox{(Number of events)} = 
  A \int_{-1}^1 dz \exp 
  \left[ \frac{-1}{2\sigma^2}
    \left( 
      M_{\tilde{W}^\pm l^\mp} 
      - M_{\tilde{W}^\pm l^\mp}^{\rm (max)} \sqrt{\frac{1+z}{2}}
    \right)^2
  \right]
  + N_{\rm BG},
\end{eqnarray}
where $A$, $M_{\tilde{W}^\pm l^\mp}^{\rm (max)}$, $\sigma$, and
$N_{\rm BG}$ are fitting parameters.
The obtained histogram is fitted in $250\ {\rm GeV}<M_{\tilde{W}^\pm l^\mp}<550\ {\rm GeV}$.
The fitted $M_{\tilde{W}^\pm l^\mp}^{\rm (max)}$ is 
estimated to be $(391\pm 6)\ {\rm GeV}$ and $(398\pm 11)\ {\rm GeV}$ for $L_T^{\rm (min)}=37.1\ {\rm cm}$ and
$44.3\ {\rm cm}$, respectively.  The uncertainty in $M_{\tilde{W}^\pm
  l^\mp}^{\rm (max)}$ as well as the difference between the best-fit
value and the underlying value (i.e., $389\ {\rm GeV}$) are typically
$\sim 10\ {\rm GeV}$ or smaller.  Thus, in the following, we assume
that the uncertainty in the measurement of the Bino mass from the
determination of the position of the endpoint is less than $\sim 10\
{\rm GeV}$, if the identification of the charged Wino track is
possible for $\tilde{W}^\pm$ going through several layers of the SCT.

\begin{figure}
 \centerline{\epsfxsize=0.6\textwidth\epsfbox{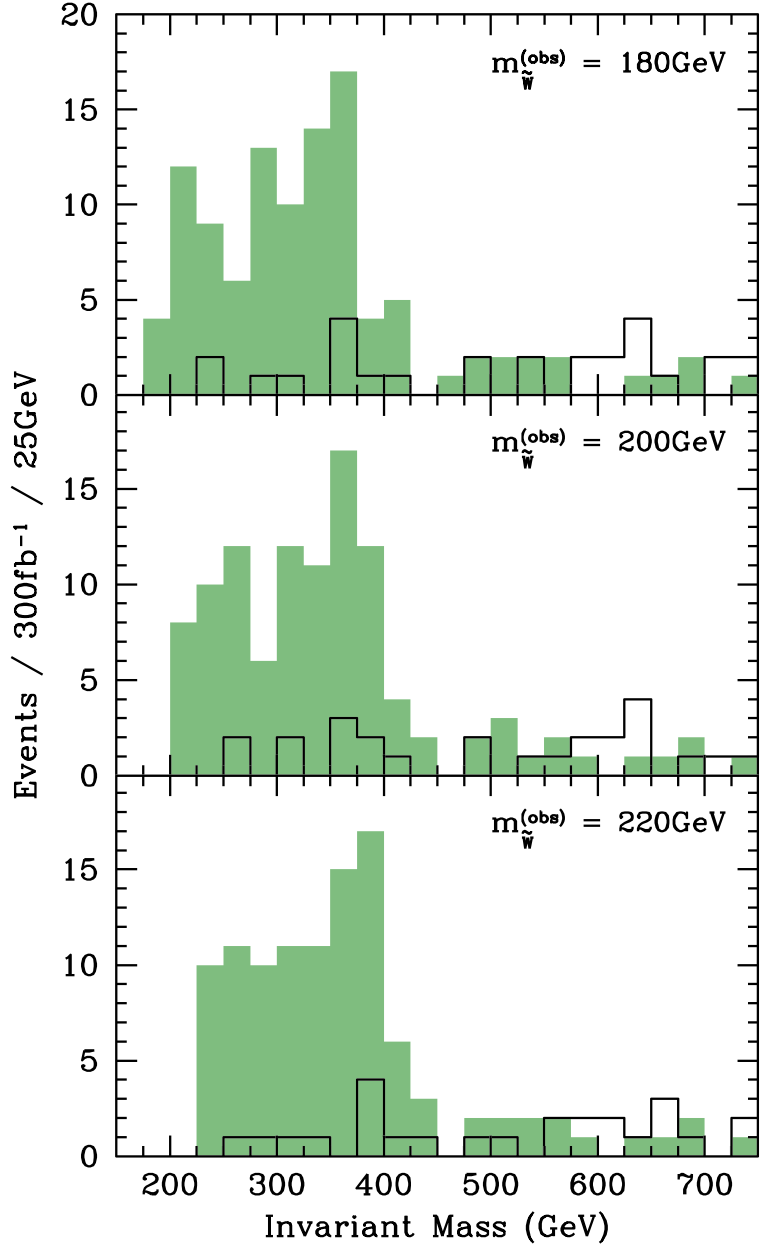}}
 \caption{\small Distribution of the opposite-sign $\tilde{W}^\pm
   l^\mp$ system (shaded histogram) and the same-sign $\tilde{W}^\pm
   l^\pm$ system (solid line) for ${\cal L}=300\ {\rm fb}^{-1}$ and
   $L_T^{\rm (min)}=44.3\ {\rm cm}$.  The postulated Wino mass is
   $180\ {\rm GeV}$, $200\ {\rm GeV}$, and $220\ {\rm GeV}$, from the
   top to the bottom.}
\label{fig:minv_mg2dep}
\end{figure}

Next, in order to see how the invariant mass distribution depends on
$m_{\tilde{W}}^{\rm (obs)}$, we calculate the invariant mass distribution
for several values of the postulated Wino mass.  The results for
$m_{\tilde{W}}^{\rm (obs)}=180\ {\rm GeV}$ and $220\ {\rm GeV}$ (as
well as for $m_{\tilde{W}}^{\rm (obs)}=200\ {\rm GeV}$) are shown in
Fig.\ \ref{fig:minv_mg2dep}, where $L_T^{\rm (min)}=44.3\ {\rm cm}$ is
used.  We can see that the endpoints depend on $m_{\tilde{W}}^{\rm
  (obs)}$.  From the figure, we expect the error of $10\ {\rm GeV}$ in
the measurement of $m_{\tilde{B}}$ from the uncertainty in the Wino
mass.  Thus, combining the uncertainty in the determination of the
endpoint, the error in the observed Bino mass is $\sim 15\ {\rm GeV}$
for the sample point we have adopted.

Here, we have adopted the integrated luminosity of $300\ {\rm fb}^{-1}$. 
We comment here that the measurement of the Bino mass may be possible with a smaller integrated luminosity
if the Wino-tracks are more efficiently produced.  
With a larger value of $c\tau$ (which is not likely in the framework of AMSB), 
or with a smaller value of the gluino mass, that can be the case.

\section{Testing the AMSB Model}

\begin{figure}
 \centerline{\epsfysize=0.75\textheight\epsfbox{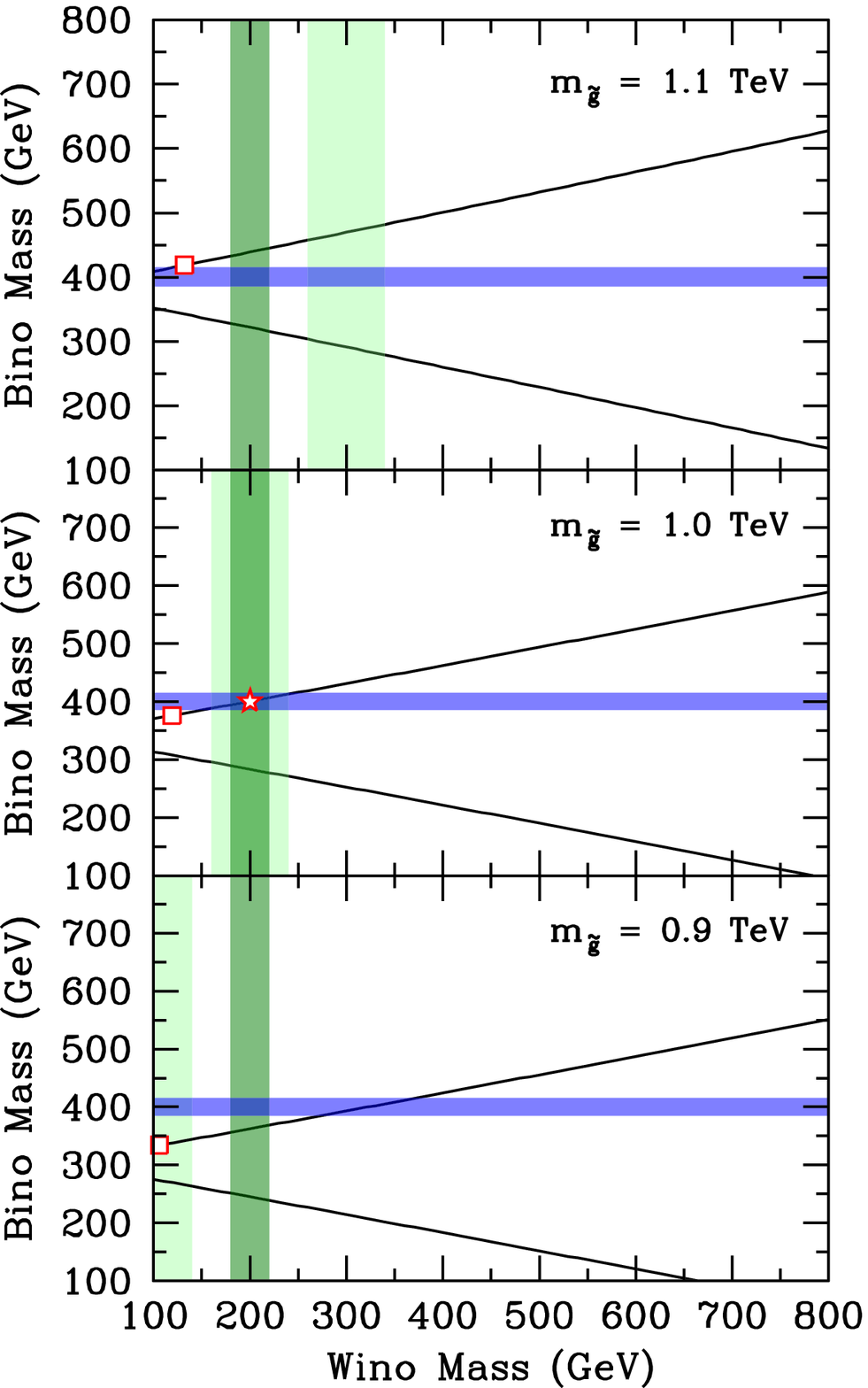}}
 \caption{\small Theoretical and experimental constraints on
   $m_{\tilde{W}}$ vs.\ $m_{\tilde{B}}$ plane for $m_{\tilde{g}}=0.9\
   {\rm TeV}$, $1\ {\rm TeV}$, and $1.1\ {\rm TeV}$ (from the bottom
   to the top).  The solid lines are upper and lower bounds on the
   Bino mass as functions of the Wino mass.  The horizontal band is
   the expected constraint on the Bino mass.  The vertical bands are
   constraints on $m_{\tilde{W}}$ (darkly shaded) and on
   $m_{\tilde{g}}- m_{\tilde{W}}$ (lightly shaded).  The star in the
   middle figure is the sample point we have used for our MC analysis,
   and the squares indicate the prediction of the minimal
   anomaly-mediated model (where gaugino masses are proportional to
   the coefficient of the beta functions of gauge coupling
   constants).}
\label{fig:m2vsm1}
\end{figure}

Finally, we discuss implication of the Bino mass determination to the
test of the AMSB model.  As indicated in Eq.\ (\ref{BoundOnM1}), the
Bino mass is constrained once the gluino and Wino masses are fixed.
In Fig.\ \ref{fig:m2vsm1}, we show the theoretical upper and lower
bounds on the Bino mass as functions of the Wino mass.  (Gluino mass
is taken to be $0.9\ {\rm TeV}$, $1\ {\rm TeV}$, and $1.1\ {\rm
  TeV}$.)  With the determination of the gaugino masses, we can see
whether they are consistent with the prediction of the AMSB model.

There are several possibilities to acquire information about gaugino masses,
as well as the Bino mass measurement.
In \cite{Asai:2007sw},
it was discussed that the mass difference
$m_{\tilde{g}}-m_{\tilde{W}}$ can be determined by studying the
distribution of the invariant mass of the di-jet emitted from the
decay process $\tilde{g}\rightarrow\tilde{W}q\bar{q}$.  With an MC
analysis, it was shown that the upper endpoint of the di-jet
invariant mass well agrees with $m_{\tilde{g}}-m_{\tilde{W}}$, and the
uncertainty in the determination of the mass difference is estimated
to be $40\ {\rm GeV}$ or so.  In addition, an information about the
gluino mass should be also obtained from the measurement of the cross
section for the process $pp\rightarrow\tilde{g}\tilde{g}$.
Furthermore, as we have mentioned, information about the Wino mass
will become available with the accuracy of $\sim 10\ \%$ once the
charged Wino is discovered in the form of the short charged tracks.
Combining those with the determination of the Bino mass, we can
perform a consistency check of the AMSB model.

The expected LHC constraints are also summarized in Fig.\
\ref{fig:m2vsm1}.  Here, we adopt the following errors in the
measurements:
\begin{eqnarray}
  \delta m_{\tilde{B}} = 15\ {\rm GeV},~~~
  \delta (m_{\tilde{g}}-m_{\tilde{W}}) = 40\ {\rm GeV},~~~
  \delta m_{\tilde{W}} = 20\ {\rm GeV}.
\end{eqnarray}
(Here, the constraint on the gluino mass from the cross section
measurement is not shown.)  Fig.\ \ref{fig:m2vsm1} shows that,
combining all the constraints, we can perform a crucial test of the
AMSB model.  We can see that the measurement of the Bino mass is very
important for such a test.  In addition, since there exist three
independent observables, all the gaugino masses are in principle
known, which can be used to determine the underlying parameters:
$m_{3/2}$, $|L|$, and $\mbox{arg}(L)$.

In our analysis, we have used leptonic decays of the $W$-boson to
determine the Bino mass.  In fact, it may be also possible to use the
hadronic decay modes if the jets from the $W$-boson can be identified,
although this may suffer from serious combinatorial backgrounds.  If
one can reduce the background, it provides a powerful method for the
Bino mass measurement.  A study of this method will be given elsewhere
\cite{AJMSY}.

{\it Acknowledgments:} The work of T.M. is supported by the
Grant-in-Aid for Scientific Research from the Ministry of Education,
Science, Sports, and Culture of Japan, No.\ 19540255.  The work of
S.S. is supported in part by JSPS Research Fellowships for Young
Scientists.  The work of T.T.Y. is supported in part by the
Grant-in-Aid for Science Research, Japan Society for the Promotion of
Science, Japan (No.\ 1940270).  This work is also supported in part by
World Premier International Center Initiative (WPI Program), MEXT,
Japan.

\end{document}